\documentclass[runningheads]{llncs}
\usepackage[T1]{fontenc}
\usepackage{graphicx}
\begin{document}
\title{Philosophical Foundations of GeoAI}
\subtitle{Exploring Sustainability, Diversity, and Bias in GeoAI and Spatial Data Science}
\author{Krzysztof Janowicz\inst{1}}
\authorrunning{K. Janowicz}
\institute{University of Vienna, Austria\\ Center for Spatial Studies, University of California, Santa Barbara\\
\email{krzysztof.janowicz@univie.ac.at}}
\maketitle              % typeset the header of the contribution
\begin{abstract}
This chapter presents some of the fundamental assumptions and principles that could form the philosophical foundation of GeoAI and spatial data science. Instead of reviewing the well-established characteristics of spatial data (analysis), including interaction, neighborhoods, and autocorrelation, the chapter highlights themes such as sustainability, bias in training data, diversity in schema knowledge, and the (potential lack of) neutrality of GeoAI systems from a unifying ethical perspective. Reflecting on our profession's ethical implications will assist us in conducting potentially disruptive research more responsibly, identifying pitfalls in designing, training, and deploying GeoAI-based systems, and developing a shared understanding of the benefits but also potential dangers of artificial intelligence and machine learning research across academic fields, all while sharing our unique (geo)spatial perspective with others.

\end{abstract}
\section{What is GeoAI?}

While GeoAI and spatial data science are relatively new fields of study, they share many of their underlying assumptions with  geography, Artificial Intelligence (AI), cognitive science, and many other disciplines while adding their own perspectives. By philosophical foundations, we mean the core principles and beliefs that underlie \textit{which} questions we ask, \textit{why} we ask them, and \textit{how} we ask them. For instance, one foundational epistemological belief underlying data science, even though rarely stated explicitly, is that knowledge can be gained through observation, a belief it shares with other empirical sciences. However, data science makes additional foundational assumptions, e.g., that (raw) data can be reused opportunistically, and that black-box methods are acceptable, i.e., that knowledge can be gained without insight. In this chapter, I will outline selected philosophical foundations of GeoAI (many also relevant to spatial data science). 

Selecting such foundations, deciding how to present them, compressing them into a few pages, and differentiating them from foundations of spatial data analysis, e.g., spatial dependence, more broadly, is challenging and not entirely objective. Hence, we will center our discussion around questions of research ethics and then highlight selected topics, such as sustainability, neutrality, and bias, from such an ethical perspective. This is for two reasons: first, too often, ethics is presented as an afterthought listed in future work sections of our papers or mentioned as an essential topic that did not make it into the curriculum of our classes. This time, we will put ethics first. Second, each topic covered here could fill hundreds of pages by itself. One way to condense these topics to just a few paragraphs is to narrow the perspective. For instance, sustainability in GeoAI could be approached from many different perspectives, e.g., purely financially. Similarly, bias (and the potential need to debias Machine Learning [ML] models) could be approached from a strictly information-theoretic perspective by noticing that bias (when defined as lack of representativeness) is essentially redundant information with less than expected information content. Finally, regarding the selection of topics, many other candidates may have been considered, e.g., privacy, reproducibility, transparency, and accountability \cite{goodchild2022white}. While these will be briefly mentioned, the main focus will be on topics that have not yet been widely considered in the GeoAI literature but should be on the mind of every researcher going forward, e.g., whether further tweaking a model is worth the environmental costs, whether the data used for evaluation is representative, or whether design decisions may affect user behavior in unintended ways.

But what is GeoAI in the first place? Just as data science is not the intersection between computer science and statistics, GeoAI should not be narrowly defined as applying AI and ML methods to use cases in the geosciences and geography.\footnote{Machine learning originated as a subfield of artificial intelligence. Hence, I will mostly use AI throughout the text and ML or AI/ML when discussing specifics, e.g., the design and training of neural networks. The term GeoML is not used in the literature (and hopefully will not be introduced).} Establishing such subfields purely by domain would increase fragmentation, thereby reducing synergies and hindering transdisciplinary research \cite{janowicz2022six}. Instead, GeoAI should incorporate spatial, temporal, and place-based (\textit{placial}) aspects into AI methods. Spatially explicit models \cite{mai2022review,li2021tobler,liu2022review,janowicz2020geoai} are such an example of successfully embedding spatial thinking into fields such as representation learning. 
Similarly, as will be discussed later, geographic classics such as the Modifiable Areal Unit Problem are used outside of our domain by the broader AI community to understand biases in their training data. Another example is the recognition that the validity of statements, e.g., in knowledge graphs, is spatially, placially, and temporally scoped. Finally, while GeoAI is by far not restricted to geography and the geosciences and has been successfully applied to downstream tasks in humanitarian relief, precision agriculture, urban planning, transportation, supply chains, climate change mitigation, and so on, most GeoAI practitioners are well trained in understanding human--environment interaction and the importance of notions such as \textit{place} that can only be defined by jointly considering physical and cognitive/societal characteristics.

\section{Ethics of GeoAI}
Ethics, as the moral compass of human behavior, is as old as philosophy. While the origins of \textit{research ethics} are challenging to trace, early work can be dated back to the 17th century and, from there on, gained traction throughout the Age of Enlightenment. Despite all its benefits, progress---be it in mathematics, engineering, medicine, or science---has never been without risks, and its benefits often favored some at the cost of others. However, it took two more centuries before the growing \textit{immediate} impact of scientific discoveries on everyday life led to widespread public recognition of the dangers of a lack of research ethics, culminating in World War II and the atomic bomb. My views on many of the challenges discussed throughout this chapter are heavily influenced by Jonas' \cite{jonas1985imperative} \textit{ethic of responsibility} in which he reformulates  Kant’s initial categorical imperative by stating that "[we should act] so that the effects of [our] action[s] are compatible with the permanence of genuine human life".

While codes and policies for research ethics vary across fields of study, agencies, countries, and so forth, most of them are based on the following five considerations:

\begin{itemize}
    \item    Who benefits from the research?
     \item   How can harm be avoided or minimized?
      \item  Who is participating, and have they given consent?
    \item    How are confidentiality and anonymity assured?
     \item   How are research outcomes disseminated?
\end{itemize}

In a narrow interpretation, the first question seeks to clarify whether the research is carried out independently, e.g., to ensure that a source funding a study does not directly or indirectly benefit from specific outcomes. In a broader sense, research should contribute to the common good of all citizens. Put differently, the first question is not merely one of integrity and objectivity but also one of justice, representatives, and the use of shared (and limited) resources.  

The second question may seem most relevant to medical research, where it aims to protect study participants from malpractices that may cause bodily harm, but also aims at minimizing socioeconomic risks. While initially primarily concerned with the individual, a broader interpretation also considers society at large. Since at least the 1960s, technology assessment (nowadays, particularly in Europe) has been an established part of research ethics. The term harm has been increasingly broadened to non-human animals and the environment in general, bringing concepts such as sustainability to the forefront.

The third consideration centers around participants and their rights. While relevant questions here are also about harm, they focus more on procedural issues. For instance, how were participants selected, and did they give informed consent? The notion of informed consent has become one of the most central concepts of research ethics well beyond medical ethics. Intuitively, consent is about informing participants about a study's objectives, procedures, and potential risks. However, informed consent also asks who can provide consent in the first place (e.g., when studies involve children), how consent has been obtained, how it can be withdrawn (e.g., by refusing further treatment), and how transparent, accessible, and understandable the goals and processes of a research study have been made. Notably, the widespread use of social media APIs throughout GeoAI research does not free the authors from some of these considerations. 

The fourth set of questions concerns confidentiality and anonymity, i.e., ensuring that data collected for a study remains secure and not traceable to individual participants. The notion of confidentiality used here extends beyond information technology measures such as encrypted, password-protected data storage. For instance, identifiable information collected about participants should not be disclosed or only via anonymization techniques. Increasingly, this includes steps to ensure \textit{privacy by design} \cite{cavoukian2009privacy} through seven principles. Key to these principles is the realization that privacy should be approached proactively instead of reactively, e.g., by minimizing the amount of data collected in the first place. Privacy by design can be regarded as a reaction to the vast body of literature demonstrating that de-identification may not efficiently protect against re-identifications attacks \cite{rocher2019estimating}, e.g., revealing location \cite{krumm2009survey,kessler2018geoprivacy,kounadi2014does}. 

Finally, the fifth consideration centers around dissemination. It asks questions such as whether research results funded by taxpayers' dollars should be hidden behind paywalls. However, ethics is about more than (free) access. Thus, the FAIR principles for data management and stewardship have also been proposed to consider issues around findability, interoperability, and reusability \cite{wilkinson2016fair}. To broaden this fifth set of questions further, reproducibility and replicability could also be regarded as ethical principles taken into account for the proper dissemination of scientific results to ensure that more people can benefit from discoveries \cite{nust2018reproducible,kedron2021reproducibility,goodchild2021replication}. 

Summing up, while most of the work on research ethics originated in fields such as medicine, cognitive science, and the social sciences, today, almost all areas of study benefit from understanding the basics of research ethics. Consequently, new branches of domain-specific ethics have been introduced to address gaps in the highly anthropocentric perspective presented before. GeoEthics is one such example \cite{peppoloni2017geoethics}. It defines principles and practices for human interaction with the environment. It introduces concepts such as geo/bio-diversity, conservation, sustainability, prevention, adaptation, and education as ethical decision-making criteria for all earth scientists. In a nutshell, GeoEthics is about establishing processes for the recognition of human responsibility for our environment.

Ethics, however, is not limited to the (direct) interaction among humans or between humans and their environment but also involves information technology, e.g., algorithms, computers, and automation more broadly. The foundations for such ethics of modern (communication) technology were laid down in the 1940s to 1980s by Norbert Wiener, Joseph Weizenbaum, Hans Jonas, James Moor, Deborah Johnson, and many others. At the core of Moor's question about what computer ethics is or should be is a dilemma about transparency that sounds all too familiar now in the 2020s: while benefits from the lightning-fast operations of computers free us from inspecting each of the millions of calculations performed per second, this lack of transparency makes us susceptible to consequences from errors, manipulation, lack of representativeness, and so on \cite{moor1985computer}. It is not surprising that Moor became interested in developing a novel concept of privacy for the digital age \cite{moor1997towards} later on. 

These and other considerations jointly form another branch of ethics, namely \textit{ethics of technology} that recognizes the social and environmental responsibilities involved in designing and utilizing computer systems and information technology \cite{jonas1985imperative}. AI ethics sits firmly within this broader branch. But what is AI ethics, and does it differ from ethics for AI? Here we will focus exclusively on our responsibilities in designing AI systems, leaving aside issues concerning behavioral norms of future (general) AI. Hagendorff \cite{hagendorff2020ethics} has compiled a recent overview of 22 ethical guidelines for AI and the aspects they cover. Most interestingly, his work provides an overview of commonalities and gaps among these frameworks. For instance, 18 out of 22 cover privacy but merely one accounts for cultural differences in the ethically aligned design of AI systems \cite{Ethicallyaligneddesign}. AI ethics generally addresses challenges arising from (a lack of) accountability, privacy, representativeness and discrimination, robustness, and explainability. This is unsurprising insofar as the answers to these ethical issues lie (at least partially) within the field of artificial intelligence and computer science more broadly. For instance, explainable AI \cite{phillips2020four} and debiasing methods \cite{bolukbasi2016man} have become widely studied areas, also in the GeoAI community \cite{LI2022101845,janowicz2018debiasing,xing2021integrating,janowicz2022geoai,papadakis2022explainable}.

However, AI ethics needs to consider broader societal implications outside of its own reach of methods to be truly impactful and to fulfill its largely positive potential. Examples of such key ethical questions are:

\begin{itemize}
    \item    How do we form societal consensus around technologies that reshape society at an unprecedented pace, e.g., regarding the future of work and education? 
    \item What are the consequences of automatic \textit{content} creation (at scale) that may be indistinguishable from human-generated content?
    \item How should we distribute the wealth created by AI?
     \item   How do we handle accountability of autonomous systems, ranging from individual autonomous cars to large parts of the financial system?
      \item  Is there a future for human judgment and decision-making in areas that require rapid response and prediction?
      \item Does intelligence require consciousness? If not, what does this mean for ethical AI?
\end{itemize}

But who, exactly, is responsible? Society, the individual data scientist, the `AI'? In their W3C Note on the responsible use of spatial data, Abhayaratna et al. \cite{w3cnote} distinguish between multiple perspectives: the developer, the user, and the regulator. Each of these roles has to contribute their part. For instance, users often all too willingly give up privacy for a bit of convenience. To give another example, developers (and scientists) should more carefully consider the minimal location precision required for a method or application to function \cite{mckenzie2022privyto}.  

In a nutshell, GeoAI ethics is an ethics of technology that recognizes the social and environmental responsibilities of developers, regulators, and users involved in designing and employing AI systems that utilize spatial, platial, and temporal data and techniques related to their data analysis.

\section{Sustainability of GeoAI}
Fueled by the emergence of foundation models \cite{bommasani2021opportunities} in 2018, progress in AI and ML has accelerated rapidly at the cost of increasingly complex models. These foundational models, such as the GPT family of language models, consist of hundreds of billions of parameters and may require terabytes of training data. Consequently, training these models produces hundreds of metric tons of carbon dioxide emissions \cite{bender2021dangers} compared to the world's annual per-capita emissions of about 4.4 tons. Going one step further and estimating the entire cradle-to-grave lifecycle of such models would quickly reveal that the environmental impact of hardware manufacturing, deployment, and decommissioning dwarfs the training and operational environmental costs \cite{gupta2022chasing,wu2022sustainable}. This is particularly concerning as it is in line with a more extensive debate about the geographic outsourcing of emissions throughout the supply chain, whereby a few counties significantly lower their (reported) footprint while, in fact, simply moving their own industry up the product (value) chain, away from manufacturing and emission-heavy stages. Put differently, the complexity and costs associated with progress in AI have grown by orders of magnitude in less than a decade, requiring new thinking about their environmental and social footprint. 

While sustainability consciousness has increased throughout the AI and ML communities, many of the proposed solutions mostly focus on selecting sites where energy consumption has a smaller carbon dioxide footprint. However, sustainability and challenges related to costs and complexity run significantly deeper. For instance, Schwartz et al. \cite{schwartz2020green} rightfully note that a Green AI should also differentiate itself from a Red AI, where increased accuracy is reached merely through sheer computational power. The authors argue that such \textit{buying} of \textit{incremental} improvement is unsustainable and raises questions of fairness, as it limits the ability to participate and compete to very few actors. 

Of course, resource consumption and resulting emissions must be put in perspective. For instance, radioactive waste is a common byproduct in hospitals, e.g., in cancer therapy. In fact, many of the most energy-intensive industry sectors, such as the chemical and construction industries, are irrevocably linked to progress and well-being. Interestingly, when changing scale from industry sectors to individual industry players, the picture becomes more complex as the list of the most energy-consuming companies is filled with big tech companies, most of them working on convenience technologies. Put differently, sustainability is not only about green(er) energy but about the inter- and intra-generational prioritization of how to utilize resources and space. Given the significant positive potential of modern-day AI for almost all areas of life, balancing its resource hunger (and other risks) with its benefits will be a significant societal challenge. 

Consequently, van Wynsberghe \cite{van2021sustainable} notes that it is essential to consider both AI for sustainability and sustainable AI, where the first is concerned with AI/ML-based methods and contributions to sustainability, e.g., environmental protection and the common good, while the second is concerned with making existing and future AI/ML systems more sustainable, e.g., by reducing emissions.

Finally, it is worth asking whether GeoAI faces additional or differently weighted sustainability challenges and whether (geo)spatial thinking offers novel perspectives on sustainability. The answer to both questions is yes. 

First, the geo-foundation models of the future may require substantially more frequent update cycles (including retraining) than other models, e.g., those needed to generate images or text. Similarly, geospatial data and models must address additional challenges related to granularity as, in theory, data can be generated at an ever-finer spatial and temporal resolution.
  
Second, thinking geographically about AI sustainability opens up new avenues to explore. For instance, one could study the relationship between regions benefiting from a certain resource-intensive model versus those regions providing these resources. Intuitively, while we all potentially benefit from research about COVID-19, as a global pandemic, it is not necessarily clear why people in Iceland should offer their (environmentally more friendly but limited) resources to models that may predominantly benefit other regions, e.g., convenience technologies in the USA. It is worth noting that this does \textit{not} imply that regions and their emissions can be seen in isolation. Finally, and to highlight yet another geographic perspective, the per-capita emissions reported here and elsewhere are global averages known to be highly skewed. Put differently, the relative burden (or, more positively, the number of people that may have benefited from the resources used) varies substantially across space (e.g., between Bangladesh and Canada), leading us to underestimate the real socio-environmental impact of very large models.

\section{Debiasing GeoAI}
Given the rapid integration of AI/ML techniques in everyday decision-making, potential biases have become an important reason for concern and, consequently, an active field of study \cite{bolukbasi2016man}. In a recent survey, Mehrabi et al. \cite{mehrabi2021survey} categorize and discuss several such biases. What these biases have in common is that they may lead to \textit{unfair}, skewed decisions. According to the authors, unfairness implies prejudice or favoritism toward some, potentially further increasing inequality. Consequently, the widespread use of AI may increase social problems such as discrimination, e.g., by widening income inequalities. The authors use the COMPAS (Correctional Offender Management Profiling for Alternative Sanctions) software as an example of a system that estimates the likelihood of recommitting crimes, all while being systematically biased against African-Americans.

Roughly speaking, bias can be introduced during three stages: from the data to the algorithm, from the algorithm to the user, and from the user interaction back to the data. For instance, intuitively, bias in the training data may cause biased models. 

To give a geographic example, according to Shankar et al. \cite{shankar2017no} 60\% of all geo-locatable images in the Open Images data set came from six countries in Europe and North America. Further, studying the geo-diversity of ImageNet, the authors reported that merely 1\% of all images were from China and 2.1\% from India. We were able to show similar coverage issues when studying potential biases in knowledge graphs \cite{janowicz2018debiasing}. Put differently, as far as commonly used data across media types is concerned, we know a lot about some parts of the world and almost nothing about others, and feeding such biased data into opaque models further exaggerates the resulting problems. One often overlooked issue is that we may overestimate (or underestimate) the accuracy of models by not considering that the difficulty of the task we are trying to address is unevenly distributed across (geographic) space. For instance, the accuracy of geoparsing systems is often reported based on unrepresentative benchmark data, leading Liu et al. to ask whether geoparsing is indeed solved or merely biased (towards specific regions to which methods have been well tailored)  \cite{liu2022geoparsing}. Other biases may arise from the well-known Modifiable Areal Unit Problem (MAUP) \cite{openshaw1984modifiable} and other sources of aggregation bias. We will address related data and schema diversity challenges separately in section \ref{diversity}. 

The next kind of biases can be introduced by the algorithm to the user, namely by causing behavioral change. One such example may be due to biased rankings in combination with the power law governing most social media. Put differently, the first ranking results are clicked over-proportionally, often followed by a quick drop. Hence, bias in ranking, e.g., of news, may alter the user's perception, e.g., of political discourse. Another bias potentially introduced at this stage is algorithmic bias, e.g., bias introduced by \textit{decisions} made during model design. We will discuss a related question, namely whether algorithms and AI are neutral, in section \ref{neutral}.

Finally, users also introduce biases into data, thereby closing the cycle back to \textit{algorithms} (here in the sense of models and their design). For instance, given that a significant part of training data across media types stems from user-generated content, changing behavior, e.g., hashtag usage over time, can introduce biases. Similarly, self-selection bias may lead systems to misestimate the result of online polls. 

Other biases can be introduced based on the mismatch between the magnitude of historical data in relation to present-day data by which models tend to make the present and future appear more like the past \cite{janowicz2018debiasing}. Such biases may affect representation learning and association rule mining and may be difficult to detect. For instance, a system may infer that given somebody is a pope, they are also male. While (potentially) controversial, this statement is true by definition. Similarly, a system may learn that if x is a US president, x must be male (given that no counterevidence exists). While currently true, this is undoubtedly not an assertion about presidents and the US we would like a system to learn.

Finally, it is worth noting that the term \textit{bias} has different meanings, and not all biases are problematic, e.g., inductive bias in ML. To make an even more abstract point, if foundation models (as basic building blocks of future AI-based systems) require algorithmic debiasing, then \textit{debiasing may itself be biased}. How transparent will this process be, and how will it account for regional, cultural, and political differences?

The examples across different bias types introduced here also serve as an important example for the argument initially made that GeoAI is not merely the application of AI/ML methods to geographic and geospatial applications. Clearly, those biases affect AI and data science more broadly while being geographic in nature and making use of well-studied concepts of spatial data analysis, such as spatial auto-correlation and the MAUP. Put differently, GeoAI contributes back to the broader AI literature. At the same time, GeoAI faces its own unique challenges related to biases.

\section{Schema and Data Diversity}\label{diversity}
Another important challenge facing GeoAI is related to the diversity of the data being processed, e.g., incorporated into training, and the diversity of the schema knowledge associated with these data \cite{janowicz2022diverse}. Data diversity can have different meanings, e.g., data coming from heterogeneous sources, data representing different perspectives, data created using different data cultures, data in various media formats, and so on. For instance, most existing ML models are not multi-modal. This, however, is changing rapidly, and future ChatGPT-like systems will handle multi-modality. Data heterogeneity remains a substantial challenge as progress on \textit{semantic} interoperability \cite{scheider2015talk} is slow. Different data cultures, e.g., governmental versus user-generated content, pose additional challenges as they have complementary strengths and weaknesses, thereby potentially requiring careful curation before being ingested into the same (Geo)AI models. Finally, multi-perspective data characterize the same phenomena but may offer complimentary or even contradictory stances. For example, the environmental footprint of nations can be assessed differently \cite{wiedmann2015material} without implying that one method is superior to others.

Taking this issue one step further brings us to schema diversity, i.e., the meaning of the domain vocabulary used may vary across space, time, and culture. While a few existing frameworks can handle contradicting assertions, e.g., about the disputed borders of the Kashmir region, schema diversity is on the terminological level, e.g., the fact that the definition of terms such as \textit{Planet}, \textit{Poverty}, \textit{Forest}, and so on are spatially, temporally, and culturally scoped. Concept drift \cite{wang2011concept}, for instance, studies the evolving nature of terms within ontologies over time (and versions). Such cases are neither well studied in the broader AI literature nor is it clear how they can be incorporated during model design, as learning is largely a monotonic process. However, given that the domains studied by GeoAI are often at the intersection between humans (society) and their environment, they will arise frequently. 

So, what exactly is the dilemma? On the one hand, data diversity is desirable for training a model for a wide range of use cases, unavoidable in Web-scale systems, e.g., knowledge graphs, and the context-dependent nature of meaning is well supported by research in linguistics and the cognitive sciences. On the other hand, however, contradicting assertions and even contradicting terminology make data curation and integration more complex and may negatively impact accuracy. Even more so, the culture-driven nature of categorization implies that category membership can change, even without substantial changes in the space of observable and engineerable (data) features \cite{janowicz2022diverse}. However, more ambiguous, changing, vague, etc. classes reduce accuracy and increase the need for training data, complex models, and so on, leading back to the discussion of sustainability \cite{janowicz2022geoai}. Note that (training) data size alone does not guarantee diversity \cite{bender2021dangers,janowicz2015data,janowicz2022diverse}. Which categorization schema we will use for geo-foundation models \cite{mai2022towards}, and how many of these models will be needed to represent regional variability remains to be seen.

\section{Is GeoAI Neutral?}\label{neutral}
"Guns don't kill, people do" is a popular slogan among parties favoring relaxed gun ownership and \textit{open carry} laws. In a nutshell, the argument states that guns are merely neutral tools, comparable to a knife or hammer, that can be used for good and evil and that humans decide to use them one way or the other. Consequently, the "mentally ill" (as the argument unfortunately goes) are the root of the problem. This slogan is often followed up by a political \textit{law and order} narrative presented as a remedy.

Similar arguments can be constructed about science and technology in general, e.g.,  by pointing out that nuclear technology was used for both the atomic bomb and for power generation and medicine. While we discussed the broader argument, counter-arguments, and technology assessment \cite{jonas1985imperative} before, the next paragraphs ask whether AI and GeoAI are neutral \cite{janowicz2022geoai}.

One key problem such discussions face is terminological confusion, e.g., terms such as ML model, architecture, system, AI, algorithm, and so forth are used interchangeably, or their meaning needs to be clarified. For instance, arguing about biases arising from selecting training data while discussing whether algorithms are neutral points to such confusion  \cite{stinson2022algorithms}. To better understand the issue, it is crucial to distinguish basic algorithms from their parameterization, e.g., via training, their deployment, their role in more sophisticated workflows such as recommender systems, and the design decisions developers take to prefer one algorithm over another. Unfortunately, these distinctions are often not explicitly made when discussing whether AI or even algorithms are neutral.

In a nutshell, computer science is about \textit{scalability} and \textit{abstraction}, but it is precisely these concepts that make discussing neutrality difficult. For instance, data structures such as queues and stacks are fundamental because they enable us to focus on the commonalities of people standing in line, whether at the mall, fuel station, or bank. In all these cases, the person (or car) getting in line first will also be served first (at least in theory). This is in stark contrast to a stack where the last item on the stack is lifted first (to get to items lower in the stack). This is true for a stack of books and kitchen plates despite all their other differences. Now, a data structure such as a stack (as an ADT) is defined by its operations, e.g., push and pop. Are the algorithms implementing these operations neutral? If so, then if algorithms (and/or AI) would not be neutral, this lack of neutrality would have to be introduced later in the process, e.g., by their combination, selection for specific downstream tasks, and so on.

To give a more geographic example, consider the shortest path in a network and whether algorithms that compute such a path are neutral \cite{janowicz2022geoai}. Dijkstra's algorithm is one such algorithm and one that follows a \textit{greedy} paradigm. While this design decision could (mistakenly) indicate a lack of neutrality, note that Dijkstra's algorithm will return the same results as other algorithms (leaving specific heuristics, nondeterministic algorithms, etc., aside). Even more, the same algorithms that are used to compute the shortest path (in fact, all paths) in a street network can also be used for many other networks, e.g., internet routing. Finally, repeated computation (again, leaving aside nondeterministic algorithms, edge/degenerated cases, issues arising from precision and parallelization, etc.) will yield the same shortest path as long as the underlying data has not changed, e.g., due to a road being blocked. So what about Dijkstra's algorithm would be non-neutral?

What does this mean for GeoAI, e.g., deep learning-based systems? Key to the success of artificial neural networks is their plasticity. In terms of their basic building blocks, and before training, these networks are essentially realized via millions and billions of multiplications (this is the scalability part) and other components such as softmax acting as an activation function. All of these single steps, such as multiplication, are by themselves neutral. Put differently, among the key reasons for the widespread success of deep learning across many downstream tasks is that the same building blocks and methods can be used, e,g., to detect buildings in remotely sensed imagery.  
This, in itself, is not a watertight argument for the neutrality of artificial neural networks or other (Geo)AI methods. Still, it shows that a meaningful discussion needs to examine where a lack of neutrality may materialize. There are several stages (some well studied) where design decisions make AI/ML based \textit{systems} non-neutral, e.g., biased. For instance, \textit{developers} make design decisions on how to combine the aforementioned basic building blocks, the data structures used, the number of hidden layers, the tuning of hyperparameters, the selection (and thereby also exclusion) of training data, the (manual) curation of data, regularization, the evaluation metrics used, the ways in which results are visualized and included further downstream to fuel rankings, predictions, and so on. Each of these steps may impact the overall performance of the system, including whether it will (implicitly) encode biases. For instance, in 2015, it was widely reported that Google's image recognition system mislabelled Black people as "gorillas" and that this might be due to a bias in the selection of training data and the (at that time) missing awareness among developers of the risks associated with (representation) bias.

This leads to two questions, namely whether these problems have been addressed sufficiently and whether there are GeoAI (or at least geo-specific) examples illustrating some issues of these implicit design decisions. As a thought example, consider the case of text-to-image models such as Stable Diffusion. In a test run of 20 prompts to generate a `Forest', none of the images depicted a forest during the winter (e.g., with snow), during the night, a rainforest, nor even basic variations in the types of trees displayed. Instead, the images produced were predominantly foggy, fall season--like images of what could best be called pine(like) forests. Prompts for `Chinese Mountains' predominantly resulted in abstract mixtures of mountains made out of fragments of the Great Wall, while `Mountains in China' did not. While these results are not representative and speak to the need for more robustness and transparency (explainability) of these early systems, they also likely show biases related to the geographic coverage of the imagery and labels used. A promising research direction for the near future will be to develop models that remain invariant under purely syntactic change. Today, for instance, embedding-based methods from the field of representation learning will yield different results for semantically similar or equivalent inputs if their syntax, e.g., structure, differs. 

To answer the initial question, today's systems are not always neutral in their representation of geographic space. This may matter as users of these systems are served very biased results.

\section{Parrots as a Chance}
Foundation (language) models have been compared to stochastic parrots \cite{bender2021dangers}. While this analogy is deeply misleading, implying that we cannot truly learn anything new from such models, it opens up new ways of thinking about their unique opportunities. Similarly to the recognition that the Modifiable Areal Unit Problem cannot be entirely avoided and should, therefore, be seen as an opportunity \cite{o2003geographic} for studying why some results change when regions are modified while others remain (relatively) stable, foundation models could be used to observe why models behave in specific ways as reflections of the underlying data and thereby society. If indeed these models encode biases based on racism, stereotypes, geography, and so on, then instead of (or better, in addition to) trying to debias these issues and thereby hide them, one could use these models to better understand the underlying phenomena at scale in ways that have never been possible in geography and the social sciences. 

For instance, the labeled (social media) data sources utilized to train text-to-image models have always been biased regarding their coverage, scenes depicted, and labels used. However, it is the ease with which we can now experiment with these models that brings the problems to the public's attention. Similarly, when learning to classify perceived safety and walkability in urban areas, we do not have to limit ourselves to arguing that these models will discriminate against certain regions and their populations, thereby further widening inequality. Instead of merely asking how to tweak such systems to yield more \textit{intended} results, we can ask which of the learned features predominantly contributed to these results and which changes to the urban areas would alter their classification. This would turn parrots into (distorted) mirrors, but we can go even further and take an observational stance by arguing that these very large models consisting of billions of parameters and trained on billions of samples are worth studying for their own sake (as they are among the largest and most complex artifacts developed by society to this day and in retrospection may provide a first glance at how communication with more general AI may look in the future). This idea aligns well with concepts such as geographic information observatories \cite{hendler2008web,janowicz2014towards,miller2017geographic}. How does the world look according to GPT-like systems, and why?

Finally, understanding why progress, e.g., in answering geographic questions, is not uniform may not only reveal problems of current model design but also point to underlying issues such as \textit{our own} tendency to favor geometry-first instead of topology-first representations \cite{feng2019relative,janowicz2022six}. For instance, in 2015, we reported that when asked for the distance between Ukraine and Russia, Google and Bing returned meaningless centroid distances instead of understanding that both countries share a border. Thus, their true distance is zero \cite{janowicz2015data}. This is still the case in 2023. Asked for the same, ChatGPT gives a substantially more sophisticated but equally meaningless answer; see Fig 1.

\begin{figure}[]
\label{fig:ur}
\centering
\includegraphics[width=1\textwidth]{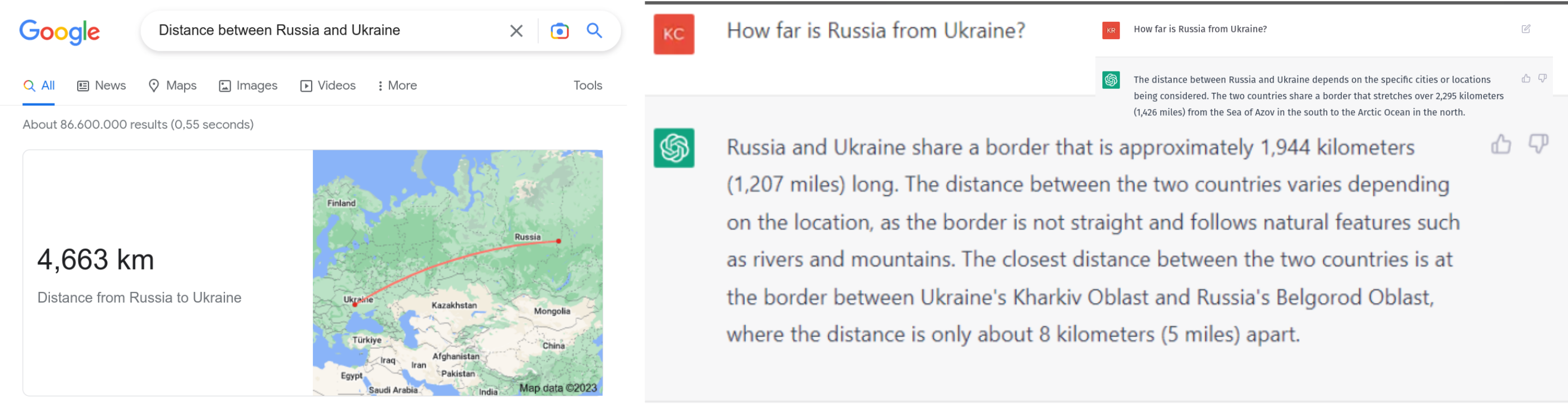}
\caption{Distance between Ukraine and Russia according to Google Search and ChatGPT. Note that ChatGPT also changes the border length across queries. The proper answer should be zero, even though ChatGPT makes a very convincing statement that at the border where both countries \textit{touch}, their distance is 5 miles.}
\end{figure}

\section{Summary and Outlook}
While GeoAI is a new and rapidly developing area that shares many of its research challenges with the broader fields of (spatial) data science, AI, geography, and the geosciences, it also offers its own questions and contributes to a broader body of knowledge. This chapter highlighted selected philosophical foundations underlying current GeoAI work from a research ethics perspective. Understanding these foundations and making them explicit is important for multiple reasons, e.g., for our community to be able to contribute to the ongoing discussion about bias and debiasing. 

However, there are also more subtle issues worth remembering. For instance, (machine) learning is based on the assumption that we can draw inferences about the present and future by studying the past. That sounds like a triviality. For today's very large language models and the resources required to train them, however, this poses many relevant challenges such as historic drag, namely that we have more data from the overall past compared to the  recent past. Not only does this imply that these models are outdated (by years), but that they may be slow to change and, therefore, may need careful curation to ensure they can keep in sync with societal change and new scientific discoveries. Similarly, as of today, these systems do not account for cultural and regional differences. Hence, it is worth exploring which theory of truth underlies their answers and our interpretation thereof (irrespective of their tendencies to hallucinate). For instance, one may speculate that systems such as ChatGPT may best be studied by following a consensus theory of truth compared to one based on coherence (which today's large language models cannot maintain).

In 2019, we provocatively asked whether "we [can] develop an artificial GIS analyst that passes a domain-specific Turing Test by 2030" \cite{janowicz2020geoai}. What if combining the first geo-foundation models of the near future with ChatGPT-like bots may get us there by 2025? Will we be able to understand and mitigate its biases? Will we be able to explain and defend our design decisions in implementing and deploying such systems? Articulating the assumptions and principles underlying GeoAI research will be a first step.

\section*{Acknowledgements}
I would like to thank Kitty Currier, Zilong Liu, Meilin Shi, Rui Zhu, Gengchen Mai, Karl Grossner, and many others for their many valuable discussions, pointers to additional literature, and critical feedback. This work was partially funded via the KnowWhereGraph project (NSF \#2033521).

 \bibliographystyle{splncs04}
 \bibliography{references}

\begin{thebibliography}{10}
\providecommand{\url}[1]{\texttt{#1}}
\providecommand{\urlprefix}{URL }
\providecommand{\doi}[1]{https://doi.org/#1}

\bibitem{w3cnote}
Abhayaratna, J., Daemen, E., Janowicz, K., Parsons, E., Smith, R., Verschoor,
  F.: The responsible use of spatial data. Tech. rep., W3C (2021)

\bibitem{bender2021dangers}
Bender, E.M., Gebru, T., McMillan-Major, A., Shmitchell, S.: On the dangers of
  stochastic parrots: Can language models be too big? In: Proceedings of the
  2021 ACM conference on fairness, accountability, and transparency. pp.
  610--623 (2021)

\bibitem{bolukbasi2016man}
Bolukbasi, T., Chang, K.W., Zou, J.Y., Saligrama, V., Kalai, A.T.: Man is to
  computer programmer as woman is to homemaker? debiasing word embeddings.
  Advances in neural information processing systems  \textbf{29} (2016)

\bibitem{bommasani2021opportunities}
Bommasani, R., Hudson, D.A., Adeli, E., Altman, R., Arora, S., von Arx, S.,
  Bernstein, M.S., Bohg, J., Bosselut, A., Brunskill, E., et~al.: On the
  opportunities and risks of foundation models. arXiv preprint arXiv:2108.07258
   (2021)

\bibitem{cavoukian2009privacy}
Cavoukian, A.: Privacy by design  (2009)

\bibitem{feng2019relative}
Feng, M., Shaw, S.L., Fang, Z., Cheng, H.: Relative space-based gis data model
  to analyze the group dynamics of moving objects. ISPRS journal of
  photogrammetry and remote sensing  \textbf{153},  74--95 (2019)

\bibitem{goodchild2022white}
Goodchild, M., Appelbaum, R., Crampton, J., Herbert, W., Janowicz, K., Kwan,
  M.P., Michael, K., Alvarez~Le{\'o}n, L., Bennett, M., Cole, D.G., et~al.: A
  white paper on locational information and the public interest  (2022)

\bibitem{goodchild2021replication}
Goodchild, M.F., Li, W.: Replication across space and time must be weak in the
  social and environmental sciences. Proceedings of the National Academy of
  Sciences  \textbf{118}(35),  e2015759118 (2021)

\bibitem{gupta2022chasing}
Gupta, U., Kim, Y.G., Lee, S., Tse, J., Lee, H.H.S., Wei, G.Y., Brooks, D., Wu,
  C.J.: Chasing carbon: The elusive environmental footprint of computing. IEEE
  Micro  \textbf{42}(4),  37--47 (2022)

\bibitem{hagendorff2020ethics}
Hagendorff, T.: The ethics of ai ethics: An evaluation of guidelines. Minds and
  machines  \textbf{30}(1),  99--120 (2020)

\bibitem{hendler2008web}
Hendler, J., Shadbolt, N., Hall, W., Berners-Lee, T., Weitzner, D.: Web
  science: an interdisciplinary approach to understanding the web.
  Communications of the ACM  \textbf{51}(7),  60--69 (2008)

\bibitem{janowicz2014towards}
Janowicz, K., Adams, B., McKenzie, G., Kauppinen, T.: Towards geographic
  information observatories. In: GIO@ GIScience. pp.~1--5 (2014)

\bibitem{janowicz2020geoai}
Janowicz, K., Gao, S., McKenzie, G., Hu, Y., Bhaduri, B.: Geoai: spatially
  explicit artificial intelligence techniques for geographic knowledge
  discovery and beyond (2020)

\bibitem{janowicz2022diverse}
Janowicz, K., Shimizu, C., Hitzler, P., Mai, G., Stephen, S., Zhu, R., Cai, L.,
  Zhou, L., Schildhauer, M., Liu, Z., et~al.: Diverse data! diverse schemata?
  Semantic Web  \textbf{13}(1), ~1--3 (2022)

\bibitem{janowicz2022geoai}
Janowicz, K., Sieber, R., Crampton, J.: Geoai, counter-ai, and human geography:
  A conversation. Dialogues in Human Geography  \textbf{12}(3),  446--458
  (2022)

\bibitem{janowicz2015data}
Janowicz, K., Van~Harmelen, F., Hendler, J.A., Hitzler, P.: Why the data train
  needs semantic rails. AI Magazine  \textbf{36}(1),  5--14 (2015)

\bibitem{janowicz2018debiasing}
Janowicz, K., Yan, B., Regalia, B., Zhu, R., Mai, G.: Debiasing knowledge
  graphs: Why female presidents are not like female popes. In: ISWC
  (P\&D/Industry/BlueSky) (2018)

\bibitem{janowicz2022six}
Janowicz, K., Zhu, R., Verstegen, J., McKenzie, G., Martins, B., Cai, L.: Six
  giscience ideas that must die. AGILE: GIScience Series  \textbf{3}, ~7 (2022)

\bibitem{jonas1985imperative}
Jonas, H.: The imperative of responsibility: In search of an ethics for the
  technological age. University of Chicago press (1985)

\bibitem{kedron2021reproducibility}
Kedron, P., Li, W., Fotheringham, S., Goodchild, M.: Reproducibility and
  replicability: opportunities and challenges for geospatial research.
  International Journal of Geographical Information Science  \textbf{35}(3),
  427--445 (2021)

\bibitem{kessler2018geoprivacy}
Ke{\ss}ler, C., McKenzie, G.: A geoprivacy manifesto. Transactions in GIS
  \textbf{22}(1),  3--19 (2018)

\bibitem{kounadi2014does}
Kounadi, O., Leitner, M.: Why does geoprivacy matter? the scientific
  publication of confidential data presented on maps. Journal of Empirical
  Research on Human Research Ethics  \textbf{9}(4),  34--45 (2014)

\bibitem{krumm2009survey}
Krumm, J.: A survey of computational location privacy. Personal and Ubiquitous
  Computing  \textbf{13},  391--399 (2009)

\bibitem{li2021tobler}
Li, W., Hsu, C.Y., Hu, M.: Tobler’s first law in geoai: A spatially explicit
  deep learning model for terrain feature detection under weak supervision.
  Annals of the American Association of Geographers  \textbf{111}(7),
  1887--1905 (2021)

\bibitem{LI2022101845}
Li, Z.: Extracting spatial effects from machine learning model using local
  interpretation method: An example of shap and xgboost. Computers, Environment
  and Urban Systems  \textbf{96},  101845 (2022)

\bibitem{liu2022review}
Liu, P., Biljecki, F.: A review of spatially-explicit geoai applications in
  urban geography. International Journal of Applied Earth Observation and
  Geoinformation  \textbf{112},  102936 (2022)

\bibitem{liu2022geoparsing}
Liu, Z., Janowicz, K., Cai, L., Zhu, R., Mai, G., Shi, M.: Geoparsing: Solved
  or biased? an evaluation of geographic biases in geoparsing. AGILE: GIScience
  Series  \textbf{3}, ~9 (2022)

\bibitem{mai2022towards}
Mai, G., Cundy, C., Choi, K., Hu, Y., Lao, N., Ermon, S.: Towards a foundation
  model for geospatial artificial intelligence (vision paper). In: Proceedings
  of the 30th International Conference on Advances in Geographic Information
  Systems. pp.~1--4 (2022)

\bibitem{mai2022review}
Mai, G., Janowicz, K., Hu, Y., Gao, S., Yan, B., Zhu, R., Cai, L., Lao, N.: A
  review of location encoding for geoai: methods and applications.
  International Journal of Geographical Information Science  \textbf{36}(4),
  639--673 (2022)

\bibitem{mckenzie2022privyto}
McKenzie, G., Romm, D., Zhang, H., Brunila, M.: Privyto: A privacy-preserving
  location-sharing platform. Transactions in GIS  \textbf{26}(4),  1703--1717
  (2022)

\bibitem{mehrabi2021survey}
Mehrabi, N., Morstatter, F., Saxena, N., Lerman, K., Galstyan, A.: A survey on
  bias and fairness in machine learning. ACM Computing Surveys (CSUR)
  \textbf{54}(6),  1--35 (2021)

\bibitem{miller2017geographic}
Miller, H.J.: Geographic information science i: Geographic information
  observatories and opportunistic giscience. Progress in Human Geography
  \textbf{41}(4),  489--500 (2017)

\bibitem{moor1985computer}
Moor, J.H.: What is computer ethics? Metaphilosophy  \textbf{16}(4),  266--275
  (1985)

\bibitem{moor1997towards}
Moor, J.H.: Towards a theory of privacy in the information age. ACM Sigcas
  Computers and Society  \textbf{27}(3),  27--32 (1997)

\bibitem{nust2018reproducible}
N{\"u}st, D., Granell, C., Hofer, B., Konkol, M., Ostermann, F.O., Sileryte,
  R., Cerutti, V.: Reproducible research and giscience: an evaluation using
  agile conference papers. PeerJ  \textbf{6},  e5072 (2018)

\bibitem{openshaw1984modifiable}
Openshaw, S.: The modifiable areal unit problem. Concepts and techniques in
  modern geography  (1984)

\bibitem{o2003geographic}
O'Sullivan, D., Unwin, D.: Geographic information analysis. John Wiley \& Sons
  (2003)

\bibitem{papadakis2022explainable}
Papadakis, E., Adams, B., Gao, S., Martins, B., Baryannis, G., Ristea, A.:
  Explainable artificial intelligence in the spatial domain (x-geoai).
  Transactions in GIS  \textbf{26}(6),  2413--2414 (2022)

\bibitem{peppoloni2017geoethics}
Peppoloni, S., Di~Capua, G.: Geoethics: ethical, social and cultural
  implications in geosciences. Annals of Geophysics  (2017)

\bibitem{phillips2020four}
Phillips, P.J., Hahn, C.A., Fontana, P.C., Broniatowski, D.A., Przybocki, M.A.:
  Four principles of explainable artificial intelligence. Gaithersburg,
  Maryland p.~18 (2020)

\bibitem{rocher2019estimating}
Rocher, L., Hendrickx, J.M., De~Montjoye, Y.A.: Estimating the success of
  re-identifications in incomplete datasets using generative models. Nature
  communications  \textbf{10}(1), ~1--9 (2019)

\bibitem{scheider2015talk}
Scheider, S., Kuhn, W.: How to talk to each other via computers: Semantic
  interoperability as conceptual imitation. Applications of Conceptual Spaces:
  The Case for Geometric Knowledge Representation pp. 97--122 (2015)

\bibitem{schwartz2020green}
Schwartz, R., Dodge, J., Smith, N.A., Etzioni, O.: {Green AI}. Communications
  of the ACM  \textbf{63}(12),  54--63 (2020)

\bibitem{shankar2017no}
Shankar, S., Halpern, Y., Breck, E., Atwood, J., Wilson, J., Sculley, D.: No
  classification without representation: Assessing geodiversity issues in open
  data sets for the developing world. arXiv preprint arXiv:1711.08536  (2017)

\bibitem{stinson2022algorithms}
Stinson, C.: Algorithms are not neutral: Bias in collaborative filtering. AI
  and Ethics  \textbf{2}(4),  763--770 (2022)

\bibitem{Ethicallyaligneddesign}
{{The IEEE Global Initiative on Ethics of Autonomous and Intelligent Systems}}:
  Ethically aligned design: A vision for prioritizing human well-being with
  autonomous and intelligent systems, first edition. IEEE (2019)

\bibitem{van2021sustainable}
Van~Wynsberghe, A.: Sustainable ai: Ai for sustainability and the
  sustainability of ai. AI and Ethics  \textbf{1}(3),  213--218 (2021)

\bibitem{wang2011concept}
Wang, S., Schlobach, S., Klein, M.: Concept drift and how to identify it.
  Journal of Web Semantics  \textbf{9}(3),  247--265 (2011)

\bibitem{wiedmann2015material}
Wiedmann, T.O., Schandl, H., Lenzen, M., Moran, D., Suh, S., West, J.,
  Kanemoto, K.: The material footprint of nations. Proceedings of the national
  academy of sciences  \textbf{112}(20),  6271--6276 (2015)

\bibitem{wilkinson2016fair}
Wilkinson, M.D., Dumontier, M., Aalbersberg, I.J., Appleton, G., Axton, M.,
  Baak, A., Blomberg, N., Boiten, J.W., da~Silva~Santos, L.B., Bourne, P.E.,
  et~al.: The fair guiding principles for scientific data management and
  stewardship. Scientific data  \textbf{3}(1), ~1--9 (2016)

\bibitem{wu2022sustainable}
Wu, C.J., Raghavendra, R., Gupta, U., Acun, B., Ardalani, N., Maeng, K., Chang,
  G., Aga, F., Huang, J., Bai, C., et~al.: Sustainable ai: Environmental
  implications, challenges and opportunities. Proceedings of Machine Learning
  and Systems  \textbf{4},  795--813 (2022)

\bibitem{xing2021integrating}
Xing, J., Sieber, R.: Integrating xai and geoai. Spatial Data Science Symposium

\end{thebibliography}

\end{document}